# UWB Radar for through Foliage Imaging using Cyclic Prefix-based OFDM

G. Ba khadher[1], A. Zidouri*[1], and A. Muqaibel[1,2]
[1]Department of Electrical Engineering
[2]Center for Communication Systems and Sensing
King Fahd University of Petroleum & Minerals
Dhahran 31261, Saudi Arabia

ghazalphds@gmail.com ,{malek,muqaibel}@kfupm.edu.sa

**Abstract**

This paper proposes the use of sufficient cyclic prefix (CP) OFDM synthetic aperture radar (SAR) for foliage penetration (FOPEN). The foliage introduces phase and amplitude fluctuation which cause the sidelobes to increase and affects the final image of the obscured targets. The wideband CP-based OFDM SAR inherently eliminates the sidelobes that arise from the interference between targets on the same range line. The integrated sidelobe level ratio (ISLR) of the CP-based OFDM signal along the range direction is lower than that of the random noise signal by 2 dB for foliage penetration application, while the peak sidelobe level ratio (PSLR) are almost the same of both of the two signals.

## 1 Introduction

Synthetic aperture radar (SAR) is used in remote sensing to provide high-resolution images of remote targets independent of weather condition and sunlight illumination in a two-dimensional spatial domain of range and azimuth. While the platform is moving, the target reflected Doppler spectrum is used to synthesize an aperture of the length of the moving path. Several types of signals have been adopted for SAR such as linear frequency modulated (LFM) [1], stepped-frequency [2] and random noise waveforms [3] [4].

Nowadays, orthogonal frequency division multiplexing (OFDM) for radar application has received a lot of attention. One of the first contributions introducing OFDM signals for radar applications was that of [5]. From then on, more research was carried out for OFDM in radar and its applications. Detection and tracking of moving target with low-grazing angle using adaptive

*: Corresponding Author: malek@kfupm.edu.sa
*ORCID:* https://orcid.org/0000-0003-2739-0208
*Scopus Author ID:* 14625802200
*Researcher ID:* B-9244-2008

OFDM radar have been studied in [6] and [7]. Closed-form expression for the compression loss due to the Doppler shift that arises from the target speed for radar coding using OFDM signal was derived in [8]. Conceptual design of OFDM as a dual system of radar and communication has been studied in [9] and [10] where the pulse diversity of this system improves its anti-detection and anti-jamming performance. A novel approach of the range profile reconstruction of OFDM radar based on the modulated symbols was developed in [11]. Adoption of multicarrier OFDM signals for SAR applications was studied in [12]. In [13], the reconstruction of the cross-range profile was developed where the azimuth components of OFDM SAR signal are separated. Then, the phase /Doppler histories of these components are estimated numerically using the least square estimation method. After that, the matched filter of the estimated phase history is used to construct the cross-range profile. The use of OFDM signal for the suppression of the range ambiguity was used by Riche et al in [13, 14, 15]. Furthermore, the authors in [16] adopted the cyclic prefix (CP) -OFDM for SAR application in order to eliminate the sidelobes that arise from the interference between targets in different range cells. However, the performance of UWB OFDM signal for SAR for through Foliage Imaging has not been investigated yet.

UWB radar for FOPEN based on the statistical physical model developed at the University of Nebraska-Lincoln (UNL) consists of two main parts; phase and amplitude fluctuations [17]. Based on the paired echo technique to analyze the pattern of the foliage, both phase and amplitude fluctuations increase the sidelobes. Though, phase fluctuation is more severe. Therefore, we propose and investigate in extension to our conference paper [19], the use of OFDM signals for radar through Foliage Imaging and particularly the use of CP-based OFDM signal. Inherently, the sidelobes that arises due to the foliage obstruction are reduced. The performance in through Foliage

Imaging is investigated and compared to the unobstructed image with random noise as a benchmark.

The remaining part of the paper is as the follows. Section II describes the geometry of OFDM SAR with sufficient CP along with the algorithm used to construct the image. The statistical physical model that represents the impact of the foliage on the transmitted signal and the received scattered signal is presented in Section III. Simulation and discussion of the performance are discussed in Section IV. Section V concludes and summarizes our recommendations for future research.

**2 OFDM SAR Signal Model**

In this work, we consider the geometry of stripmap broadside SAR for through Foliage Imaging in Fig.1. An airplane is moving parallel to the azimuth direction with an instantaneous coordinate $(0, y_p(\eta), H_p)$ where the azimuth time is $\eta$ and $H_p$ is the altitude of the radar platform. $T_a$ is defined as the time extent along the flight over which the target on the ground lies in the antenna beam; synthetic aperture time.

Consider that at the transmitter side, an OFDM signal with $N$ subcarriers and bandwidth of $B$ Hz is to be transmitted, and let $\boldsymbol{X} = [X_0, X_1, \ldots, X_{N-1}]$ be the population of the symbols in the frequency domain. Then the discrete time domain-OFDM signal is obtained by the Inverse Fast Fourier Transform (IFFT) of the vector $\boldsymbol{X}$. The can write the OFDM signal as:

$$s(t) = \frac{1}{\sqrt{N}} \sum_{k=0}^{N-1} X_k \exp\left\{\frac{j2\pi kt}{T}\right\}, t \in [0, T + T_{GI}] \qquad (1)$$

where $t$ is the length of the OFDM signal that consists of two parts, the time duration of the OFDM signal without the CP is $T$ and the length of the CP is $T_{GI}$.

For $T = NT_s$ and $T_{GI} = (M-1)T_s$ where $T_s = \frac{1}{B}$ is the sampling frequency, $M$ is the number of range cells and $N$ is number of subcarriers. After sampling at $t = iT_s$, we can write (1) as

$$s_i = s(iT_s) = \frac{1}{\sqrt{N}} \sum_{k=0}^{N-1} X_k \exp\left\{\frac{j2\pi ki}{N}\right\}, i = 0, 1, \ldots, N + M - 2. \qquad (2)$$

At the receiver side and after demodulation to baseband, the signal complex envelope from fixed-point object in the $m^{th}$ range cell can be modeled in terms of slow time $\eta$ and fast time $t$

$$z_m(t, \eta) = \sigma_m \, \varepsilon_a(\eta) \exp\left\{-j4\pi f_c \frac{R_m(\eta)}{c}\right\}$$

$$\times \frac{1}{\sqrt{N}} \sum_{k=0}^{N-1} X_k \exp\left\{\frac{j2\pi k}{T}\left[t - \frac{2R_m(\eta)}{c}\right]\right\} + w(t, \eta), \qquad (3)$$

$$t \in \left[\frac{2R_m(\eta)}{c}, \frac{2R_m(\eta)}{c} + T + T_{GI}\right]$$

where $\varepsilon_a(\eta) = p_a^2(\theta(\eta)) \approx \operatorname{sinc}\left(\frac{L_a \theta}{\lambda}\right)^2$ is the azimuth beam that determines the strength of the received signal along the azimuth direction, $\theta$ is the angle measured from the boresight in the slant range plane, and $L_a$ is the antenna effective length. $\sigma_m$ represents the radar cross section (RCS) coefficient from the target in the $m^{th}$ range cell within the footprint of the radar beam, $c$ is the speed of light, and $w(t, \eta)$ represents the noise. The slant range $R_m(\eta)$ between the radar and the target in the $m^{th}$ range cell with the coordinate $(x_m, y_m, 0)$ may be written as

$$R_m(\eta) = \sqrt{x_m^2 + H_P^2 + v_p^2 \eta^2} \qquad (4)$$

where $v_p$ is the radar platform effective velocity. The complex envelope of the received signal from every range cell within the swath width can be given as

$$z(t, \eta) = \sum_m z_m(t, \eta). \qquad (5)$$

We may convert the received data in (5) to the discrete time linear convolution of the transmitted sequence $s_i$ in equation (2) with the weighting radar cross section coefficients $g_m$ which may be written as

$$z_i = \sum_{m=0}^{M-1} g_m s_{i-m} + w_i, i = 0, 1, \dots, N + 2M - 3 \qquad (6)$$

where

$$g_m = \sigma_m \, \varepsilon_a(\eta) exp\left\{-j4\pi f_c \frac{R_m(\eta)}{c}\right\} \qquad (7)$$

From the received signal of equation (6), the first and the last $M - 1$ samples are removed. Then, we obtain the following:

$$z_n = \sum_{m=0}^{M-1} g_m s_{n-m} + w_n, n = M - 1, M, \dots, N + M - 2 \qquad (8)$$

Then, the received signal can be expressed as $\mathbf{z} = [z_{M-1}, z_M, \cdots, z_{N+M-2}]$

The OFDM demodulator performs an FFT (Fast Fourier Transform) on the vector $\mathbf{z}$.

$$Z_k = \frac{1}{\sqrt{N}} \sum_{n=0}^{N-1} z_{n+M-1} exp\left\{\frac{-j2\pi kn}{N + M - 1}\right\}, k = 0, 1, \dots, N - 1 \qquad (9)$$

The aforementioned $Z_k$ can be expressed as

$$Z_k = G_k X_k + W_k, k = 0,1, \dots, N-1. \tag{10}$$

where $X_k$ are the symbols transmitted, $W_K$ is the FFT of the noise, and

$$G_k = \sum_{m=0}^{M-1} g_m exp\left\{\frac{-j2\pi mk}{N}\right\}, k = 0,1, \dots, N-1. \tag{11}$$

Therefore, the estimate of $G_k$ is

$$\hat{G}_k = \frac{Z_k}{X_k} = G_k + \frac{W_k}{X_k}, k = 0,1, \dots, N-1. \tag{12}$$

The vector $\boldsymbol{G} = [G_0, G_1, \dots, G_{N-1}]^T$ in (12) is the N-point FFT of $\sqrt{N}\psi$, where $\psi$ is the weighting Radar Cross Section (RCS) coefficient vector

$$\psi = [g_0, g_1, \dots, g_{M-1}, 0, \dots, 0] \tag{13}$$

The estimation of the weighting RCS coefficient $g_m$ can be obtained by performing N inverse FFT point on the vector $\hat{\boldsymbol{G}} = [\hat{G}_0, \hat{G}_1, \dots, \hat{G}_{N-1}]^T$.

$$\hat{g}_m = \frac{1}{\sqrt{N}} \sum_{k=0}^{N-1} \hat{G}_k \, exp\left\{\frac{j2\pi mk}{N}\right\}, m = 0, \dots, M-1. \tag{14}$$

Afterwards, the estimation of the weighting RCS coefficients of the $M$ cells along the range direction may be written as

$$\hat{g}_m = \sqrt{N}\, g_m + \widetilde{w}'_m, m = 0, \dots, M-1. \tag{15}$$

where $\widetilde{w}'_m$ represents noise which has the same variance as in (12).

When the weighting RCS coefficients $g_m$ are determined, the RCS coefficients $\sigma_m$ can be obtained from (7) and vice versa as follows

$$\hat{\sigma}_m = \hat{g}_m exp\left\{j4\pi f_c \frac{R_m(\eta)}{c}\right\} \tag{16}$$

The focusing in the azimuth direction is like the conventional stripmap SAR [19] as shown in Fig. 2(a). The azimuth compression and the range cell migration correction (RCMC) are implemented in all the swath range, using fixed value for the reference range cell, $R_C$, for computational efficiency. For comparison, we consider the random noise signal [3], a band-limited wide-sense stationary (WSS) Gaussian process with zero mean and variance $\sigma^2$ which is given as

$$s(t) = s_I(t) \cos(2\pi f_0 t) - s_Q(t)\sin(2\pi f_0 t) \qquad (17)$$

where $s_I(t)$ and $s_Q(t)$ are Gaussian random processes, and $f_0$ is the central frequency. The reconstruction of SAR image using range Doppler algorithm is shown in Fig. 2, which consists of two parts. Fig. 2(a) presents the processing of CP-based OFDM signal and starts by the removal of the cyclic prefix followed by the transformation into the frequency domain using FFT. Then, the estimation of the weighting radar cross section coefficient is performed by dividing the received information by the transmitted and transformation into the time domain with the help of IFFT as in (14). The data along the azimuth direction is transformed into the frequency domain. Then the range cell migration correction (RCMC) and the azimuth compression are implemented. Fig. 2 (b) illustrates the processing of random noise signal and starts by the correlation between the transmitted signal and the range time radar data. The difference between the processing of two signals is in the range reconstruction, while the RCMC and the azimuth compression are similar.

**3 Model of Foliage Penetration**

In order to introduce the effect of the foliage obscuration on the transmitted and the received signal for SAR FOPEN, the FOPEN radar imaging need to be presented in the frequency domain as shown in Fig. 3. $\boldsymbol{F_T}(\boldsymbol{\omega}, \boldsymbol{\eta}, \boldsymbol{\gamma_g})$ and $\boldsymbol{F_R}(\boldsymbol{\omega}, \boldsymbol{\eta}, \boldsymbol{\gamma_g})$ are the foliage propagation characteristics for the transmitted signal and the received target scatterer respectively. $\boldsymbol{G_F}(\boldsymbol{\omega}, \boldsymbol{\eta})$ represents the received

obscured signal. The foliage obscured target range profile, in the frequency domain, may be modeled as

$$G_F(\omega,\eta,\gamma_g) = S(\omega,\eta,\gamma_g)F_T(\omega,\eta,\gamma_g)G(\omega,\eta)F_R(\omega,\eta,\gamma_g)$$
$$= G_s(\omega,\eta,\gamma_g)F(\omega,\eta,\gamma_g) \qquad (18)$$

where $F(\omega,\eta,\gamma_g) = F_T(\omega,\eta,\gamma_g)F_R(\omega,\eta,\gamma_g)$, is the frequency and the flight path dependent two-way foliage transmission at specific grazing angle and polarization.

The two-way foliage transmission model according to [17] can be described by a transfer function, which has, nonlinear amplitude characteristic and nonlinear phase characteristic. This transfer function at specific polarization can be given as

$$F(\omega,\eta,\gamma_g) = A(\omega,\eta,\gamma_g)exp\,[j\Phi(\omega,\eta,\gamma_g)] \qquad (19)$$

where $A(\omega,\eta,\gamma_g)$ represents the nonlinear amplitude characteristic and $\Phi(\omega,\eta,\gamma_g)$ indicates the nonlinear phase characteristic. Both the amplitude $A$ and the phase $\Phi$ are functions of the radar frequency, the azimuth path and the grazing angle $\gamma_g$.

### 3.1. Amplitude characteristics

The mean attenuation and the amplitude fluctuation constitute the amplitude characteristic which can be represented as

$$A(\omega,\eta,\gamma_g) = A_0(\omega,\gamma_g)[1 + \delta_A(\omega,\eta,\gamma_g)] \qquad (20)$$

where $A_0(\omega,\gamma_g)$ represents the mean attenuation and $\delta_A(\omega,\eta,\gamma_g)$ is the normalized amplitude fluctuation.

According to [18], the mean attenuation of the foliage can be written as

$$A_0(\omega, \gamma_g) = \beta f^\alpha (\sin 45° / \sin \gamma_g) \tag{21}$$

where $A_0(\omega, \gamma_g)$ is in dB, $f$ is the radar center frequency, $\gamma_g$ is the grazing angle to the local clutter patch, and $\alpha$ and $\beta$ are two constants. We summarize the used values of $\alpha$ and $\beta$ in Table 1.

The normalized amplitude fluctuation $\delta_A(\omega, \eta, \gamma_g)$ consists of two components, the grazing angle and frequency dependent $\delta_\omega(\omega, \gamma_g)$ and the flight path dependent amplitude fluctuation $\delta_\eta(\eta)$ which can be expressed as follows

$$\delta_A(\omega, \eta, \gamma_g) = \delta_\omega(\omega, \gamma_g)\delta_\eta(\eta) \tag{22}$$

The grazing angle and the frequency dependence are modeled as Gamma probability density random process as,

$$p(x, a, b) = \frac{1}{b^a \Gamma(a)} x^{a-1} e^{-x/b} \tag{23}$$

where $a$ and $b$ are two constants to be determined by the mean attenuation and the variance statistics of the measured amplitude fluctuation. The mean and the variance of the Gamma distribution are $\mu = ab$ and $\sigma^2 = ab^2$, respectively.

The flight path dependent amplitude fluctuation $\delta_\eta(\eta)$ is modeled as

$$\delta_\eta(\eta) = \exp(\eta_H(\Delta\eta)) \tag{24}$$

where $\eta_H(\Delta\eta)$ represents the fractional Brownian motion (fBm) random process which has two parameters; the Hurst exponent H≃ 0.4 for vegetation cover [17] and $\Delta\eta$ which is related to the synthetic aperture size or the length of the flight path.

### 3.2. Phase Characteristics

The phase characteristics consist of the phase fluctuation which may be expressed as

$$\Phi(\omega,\eta,\gamma_g) = \delta_\phi(\omega,\eta,\gamma_g) \tag{25}$$

The phase fluctuation can be derived from the amplitude fluctuation with the assumption that the phase of the incoherent field is uniformly distributed from $-\pi$ to $\pi$ as the following:

$$\delta_\phi(\omega,\eta,\gamma_g) = \tan^{-1}\left[\frac{\delta_A(\omega,\eta,\gamma_g)\sin(\psi)}{1+\delta_A(\omega,\eta,\gamma_g)\cos(\psi)}\right] \tag{26}$$

where $\psi$ has a uniform density over $[-\pi,\pi]$.

In order to introduce the effect of the foliage obstruction into the focused image of SAR FOPEN system, the received raw radar data along the range direction, is transformed into the frequency domain and multiplied by the foliage obscured frequency domain signatures. This is illustrated in Fig. 4. This can be illustrated mathematically as the Following

$$G_k(\omega,\eta,\gamma_g) = Z_k \times F_k(\omega,\eta,\gamma_g) + w_k, k = 0,1,\ldots,N+2M-3 \tag{27}$$

where $Z_k = \frac{1}{\sqrt{N}}\sum_{i=0}^{N+2M-3} z_i \exp\left\{\frac{-j2\pi ki}{N}\right\}$ which is the FFT of the received vector $z$ in (6). After that IFFT is applied along the range direction to get the time-domain version of the received echo. The reconstruction processing of SAR FOPEN is illustrated in Fig.4, which is similar to Fig. 2 apart from the incorporation of the foliage effect through the transformation of the data along the range time direction into the range frequency domain in order to multiply it with the foliage obscuration effect and going back into the time domain. The remaining steps of the processing are similar to that in Fig. 2.

## 4 Simulation and Performance Evaluation

MATLAB simulation was carried out to investigate the UWB Cyclic Prefix based OFDM SAR and UWB random noise SAR for the application of FOPEN using the following parameters. The Bandwidth is $B = 4$ GHz, and the carrier frequency $f_s = 9$ GHz. The time to synthesize the aperture is $T_a = 1$ s, the effective speed of the moving radar $v_p = 150$ m/s, the center of the slant range swath is $R_c = 5\sqrt{2}$ km, the height of the platform is $H_p = 5$ km, and the number of cells along the range direction is $M = 192$. The duration of OFDM signal without CP is $T = 256$ ns, the number of OFDM subcarriers is $N = 1024$, the length of the CP is $CP_{length} = 191$ and the duration of the CP is $T_{GI} = 47.75$ ns. Therefore, the length of the CP-based OFDM is $T_0 = 303.75$ ns. The time duration of the random noise signal is also the same as for CP-based OFDM signal. The population of the symbols in the frequency domain over the UWB CP-based OFDM SAR's subcarriers is considered to be vectors of the binary pseudorandom noise sequence corresponding to values of $-1$ and $1$. We are considering scenarios for both single point target and extended target.

First, a single point target is located at the center of the swath width. To be able to assess the impact of foliage, polarization, and type of signal used. Four images are reconstructed representing all possible combination (HH/VV polarization, random noise signal/CP-based OFDM signal). The first two images are reconstructed assuming no-foliage. HH polarization with CP-based OFDM and random noise are illustrated in Fig. 5(a), Fig. 5 (b), respectively. While the cases of VV polarization are the same as the HH polarization. It is clear that the CP-based OFDM signal has superior performance compared to random noise signal.

Furthermore, two similar images are reconstructed assuming foliage with HH polarization as in Fig. 6. While for the VV polarization are almost the same as the HH polarization. It is evident that effect of the foliage blurred the images of SAR system. The normalized range profiles of the UWB CP-based OFDM signal and UWB random noise signals with and without application of FOPEN are illustrated in Fig. 7. It can be noticed that the sidelobes for UWB CP-based OFDM signal are lower than that of the UWB random noise's sidelobes. The normalized azimuth profiles of the point spread function of the two signals are illustrated in Fig. 8. It can be seen that the azimuth profiles with or without the application of FOPEN are similar for both of the two signals.

For quantitative evaluation, two measures are used to investigate the performance of CP-based OFDM signal for FOPEN compared to the random noise signal for the SAR FOPEN which is the integrated sidelobe level ratio (ISLR) and the peak sidelobe level ratio (PSLR) [18].

ISLR is defined as the ratio of the total sidelobes on both sides of the main lobe to the main lobe which is expressed in decibels as:

$$ISLR = 10\log \left(\frac{power\ integrated\ over\ sidelobes}{total\ main\ lobe\ power}\right) \qquad (28)$$

PSLR is defined as the ratio between the height of the largest side lobe and the height of the main lobe which in decibels is given as:

$$PSLR = 10\log \left(\frac{power\ integrated\ over\ sidelobes}{total\ main\ lobe\ power}\right) \qquad (29)$$

The results of these metrics for the two signals are illustrated in Tables 3 and 4. The ISLR and the PSLR for both of the two signals along the azimuth direction are the same. However, The ISLR of the CP-based OFDM signal is about 3.2 dB lower than that of the random noise signal, while the PSLR of the CP-based OFDM signal is also about 2.6 dB lower than that of the random noise signal along the range direction.

Similarly, for the application of FOPEN, The ISLR and the PSLR for both of the two signals along the azimuth direction are the same. On the other hand, the ISLR of the CP-based OFDM signal is about 2.3 dB lower than that of the random noise signal, while the PSLR for both of two signals are almost the same along the range direction.

We then further extend the work by considering an extended target with the shape of a tank by the arrangement of few single point targets. The results are illustrated in Fig. 9, and Fig. 10 with the possible combination of no foliage and foliage respectively. The cases of VV polarization are the same as the HH polarization with no foliage, while with no foliage are almost the same. It can be clearly noticed that the CP-based OFDM signal outperforms the random noise signal.

## 5 Conclusions

In this paper, we have proposed CP-based OFDM signal for the application of FOPEN imaging. We have investigated the performance compared with the random noise signal as a benchmark. It can be seen that all the two radar systems are affected by the foliage. However, the CP-based OFDM signal for the FOPEN application is better than the random noise radar because the fluctuation of the sidelobes along the range direction is lower than the latter. Therefore, our results corroborate the proposition of UWB Cyclic Prefix-based OFDM radar to be used for FOPEN SAR.

**Acknowledgments**

The authors would like to acknowledge the support by King Fahd University of Petroleum & Minerals (KFUPM).

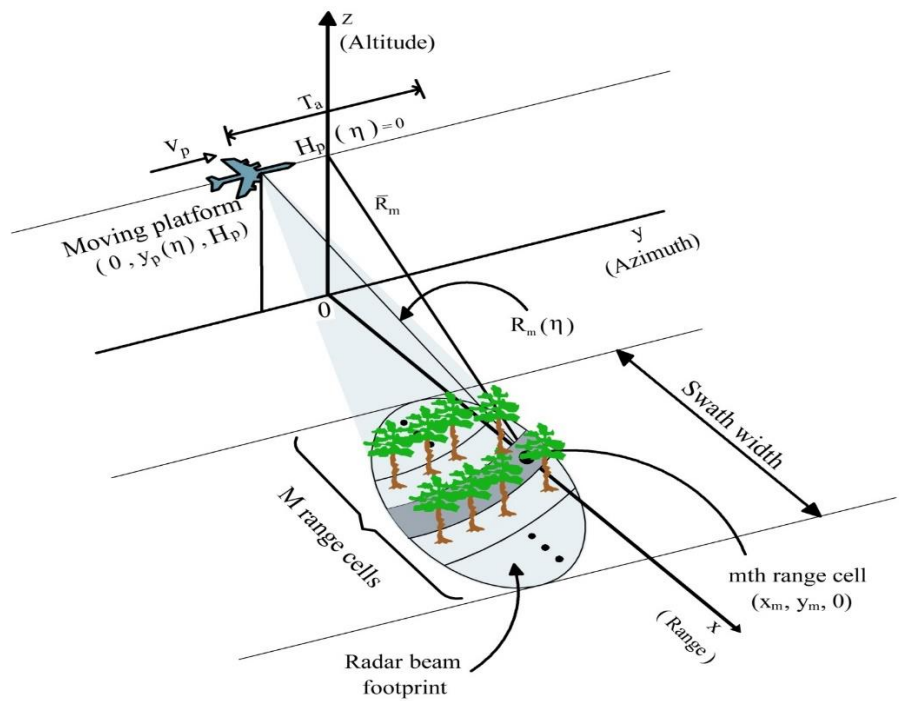

*Fig. 1. Geometry of stripmap SAR FOPEN*

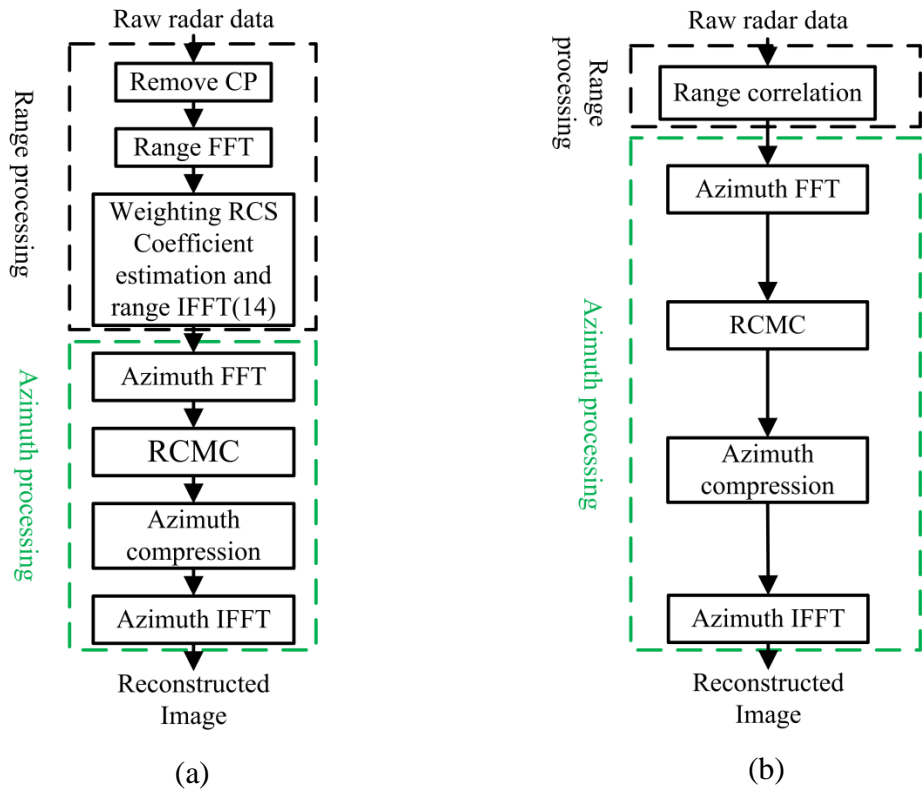

*Fig. 2. Block diagram of SAR imaging processing. (a) CP-based OFDM SAR. (b) Random noise SAR*

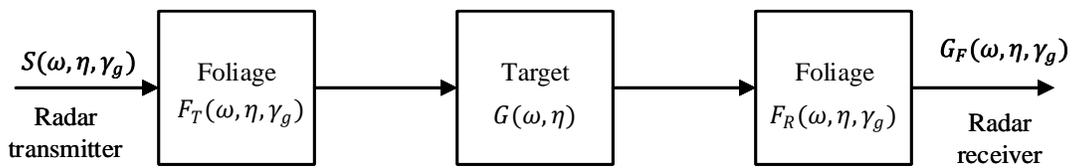

*Fig. 3 FOPEN radar imaging*

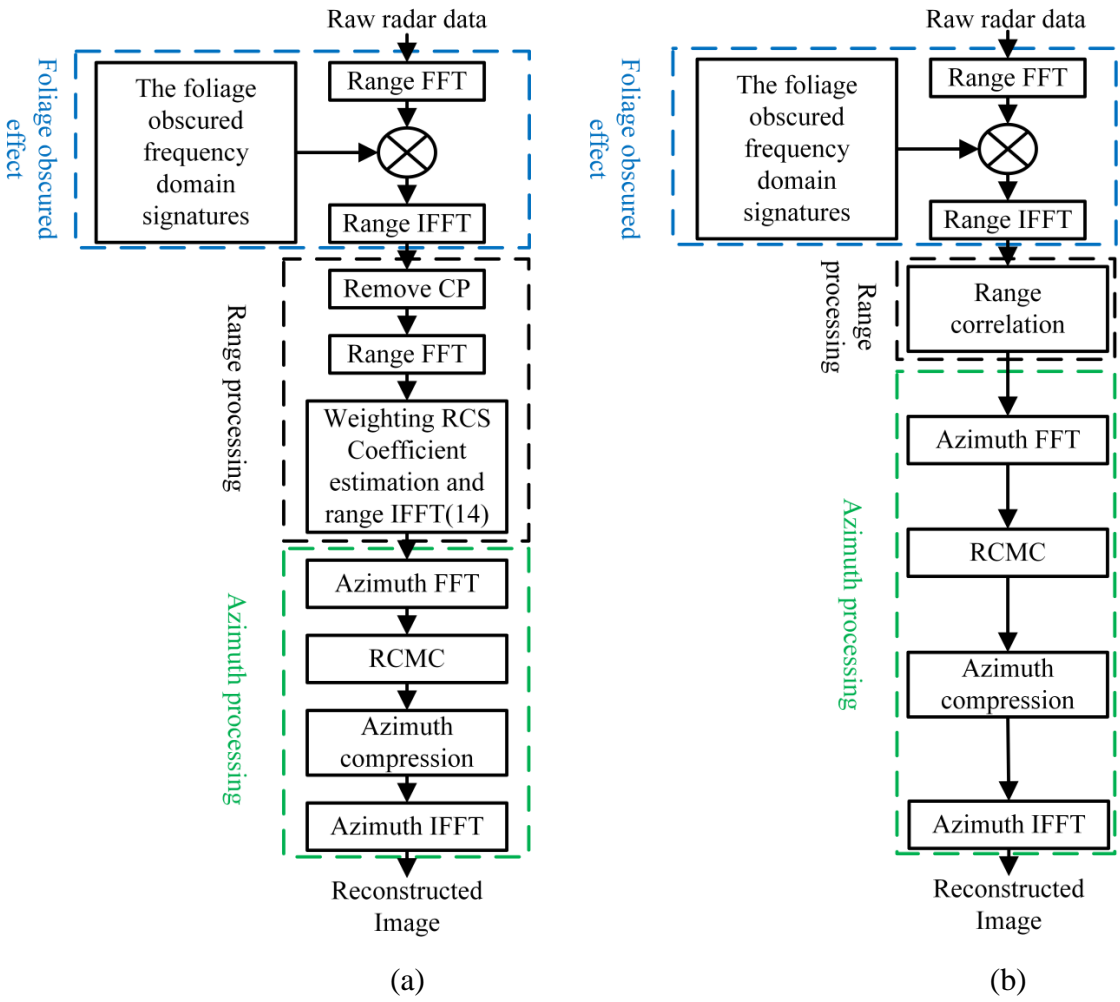

Fig. 4. Block diagram of UWB SAR imaging process for FOPEN. (a) CP-OFDM signal. (b) Random noise signal

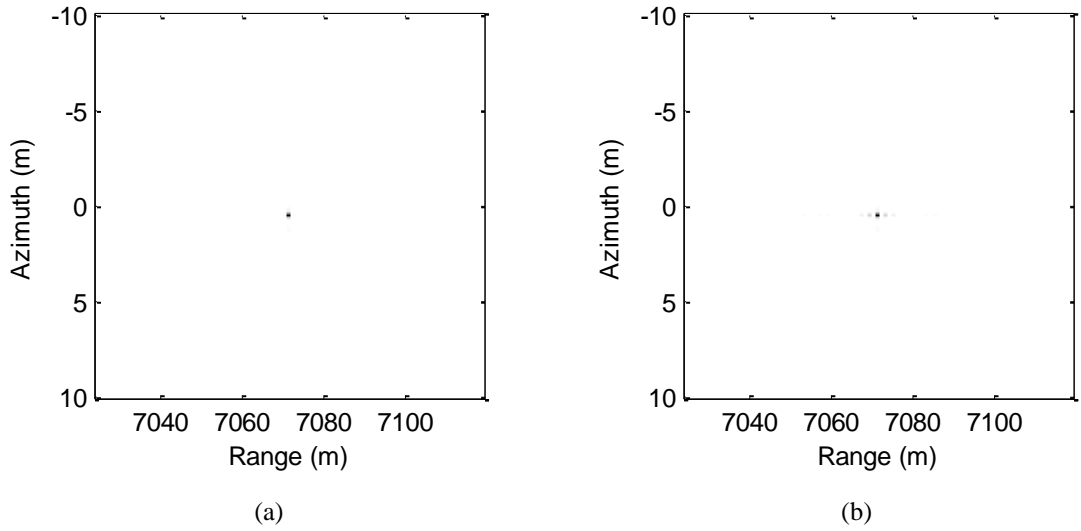

*Fig. 5. Single point target Imaging results for UWB SAR. (a) CP-based OFDM SAR, HH. (b) Random noise SAR, HH.*

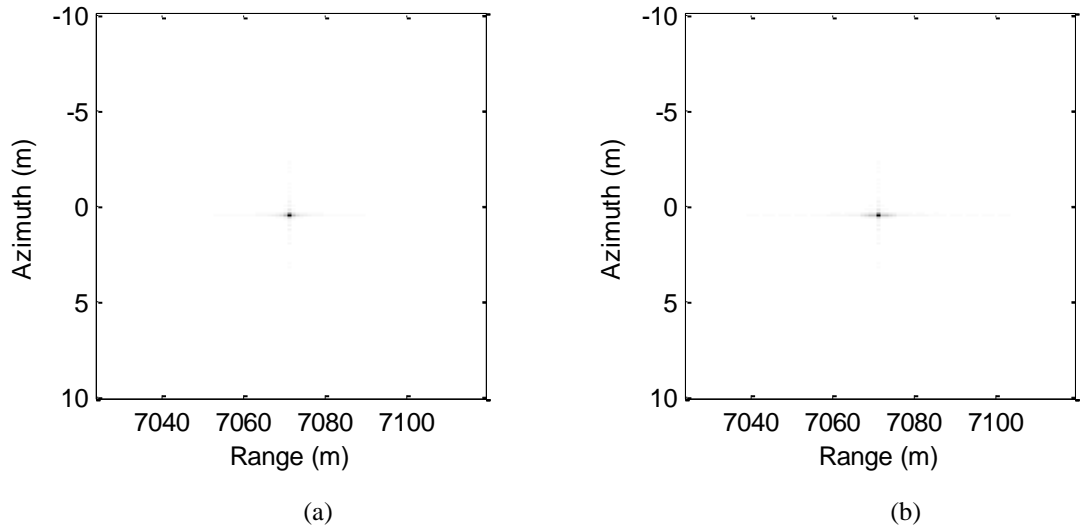

*Fig. 6. Single point target Imaging results of UWB SAR FOPEN. (a) CP-based OFDM SAR FOPEN, HH. (b) Random noise SAR FOPEN, HH.*

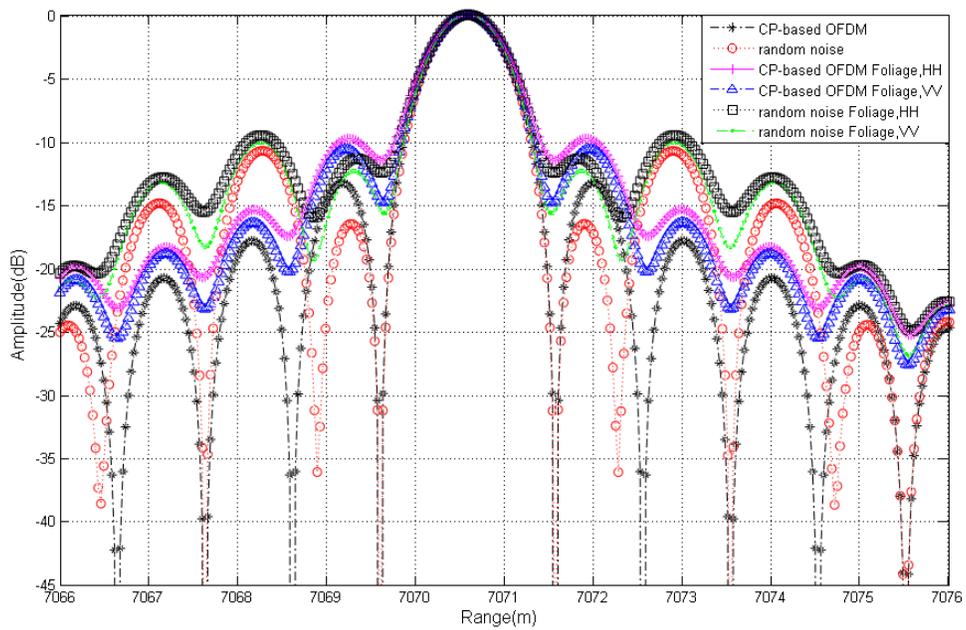

*Fig. 7. Spread function Range profiles*

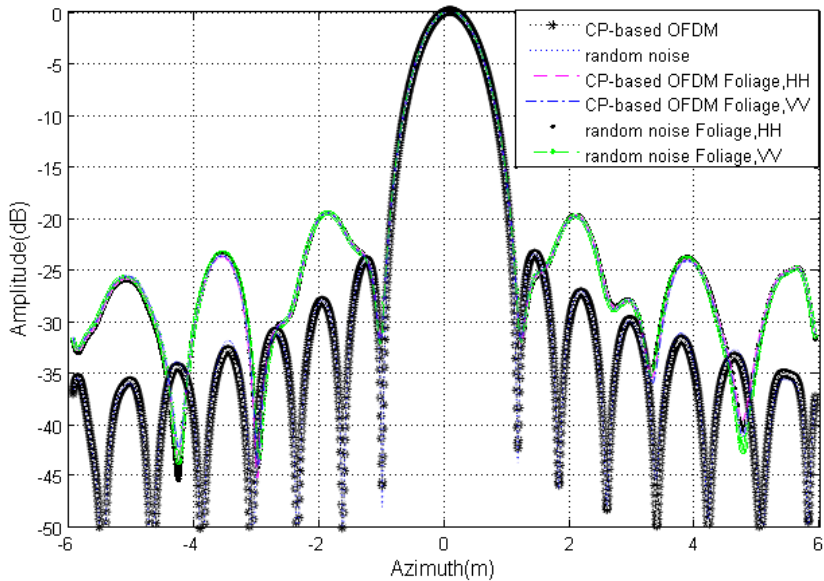

*Fig. 8. . Spread function Azimuth profiles*

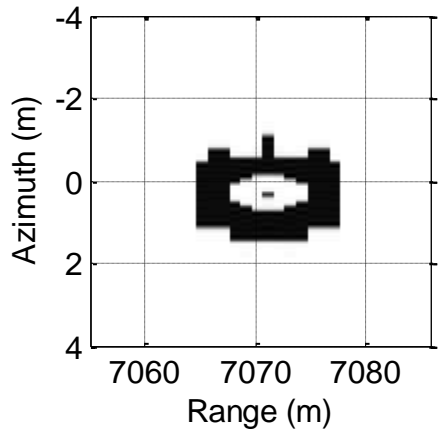 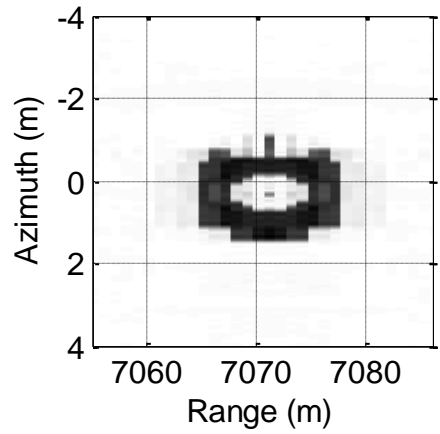

(a)                                                (b)

*Fig. 9. Extended target Imaging results of UWB SAR. (a) CP-based OFDM SAR, HH. (b) Random noise SAR, HH.*

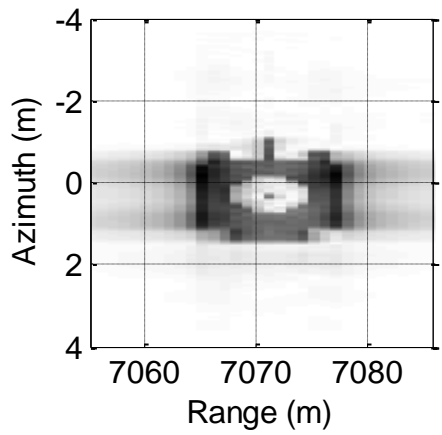 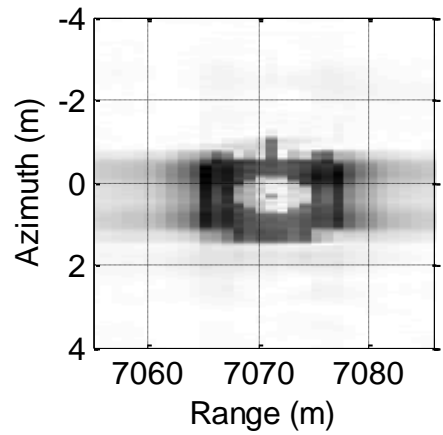

(a)                                                (b)

*Fig. 10. Extended target Imaging results of UWB SAR FOPEN. (a) CP-based OFDM SAR, HH. (b) Random noise SAR, HH.*

Table 1 Model parameters for Foliage attenuation

| Mean attenuation Model | | |
|---|---|---|
| Polarization | $\alpha$ | $\beta$ |
| HH | 0.79 | 0.05 |
| VV | 0.5 | 0.45 |

Table 2 Results of image quality metrics for UWB CP-based OFDM and UWB random noise SAR

| Method | | CP-OFDM, HH | Random noise, HH | CP-OFDM, VV | Random noise, VV |
|---|---|---|---|---|---|
| $ISLR_{dB}$ | Range | −9.68 | −6.47 | −9.68 | −6.47 |
| | Azimuth | −21.78 | −21.71 | −21.78 | −21.71 |
| $PSLR_{dB}$ | Range | −13.26 | −10.71 | −13.26 | −10.71 |
| | Azimuth | −23.49 | −23.49 | −23.49 | −23.49 |

Table 3 Results of image quality metrics for UWB CP-based OFDM and UWB random noise SAR for FOPEN

| Method | | CP-OFDM, HH | Random noise, HH | CP-OFDM, VV | Random noise, VV |
|---|---|---|---|---|---|
| $ISLR_{dB}$ | Range | −5.63 | −3.37 | −6.49 | −4.18 |
| | Azimuth | −15.26 | −15.24 | −15.29 | −15.25 |
| $PSLR_{dB}$ | Range | −9.76 | −9.49 | −10.57 | −10.06 |
| | Azimuth | −19.52 | −19.52 | −19.54 | −19.52 |